\documentclass[12pt]{article}
\usepackage{epsfig,amsmath,amsfonts,amssymb,amstext,afterpage,psfrag,slashbox}
\def\spose#1{\hbox to 0pt{#1\hss}}
\def\lsim{\mathrel{\spose{\lower 3pt\hbox{$\mathchar"218$}}
 \raise 2.0pt\hbox{$\mathchar"13C$}}}
\def\gsim{\mathrel{\spose{\lower 3pt\hbox{$\mathchar"218$}}
 \raise 2.0pt\hbox{$\mathchar"13E$}}}

\catcode`@=11
\def\@citex[#1]#2{%
  \if@filesw\immediate\write\@auxout{\string\citation{#2}}\fi
  \def\@citea{}\@cite{\@for\@citeb:=#2\do
    {\@citea\def\@citea{,\penalty\@m}\@ifundefined
      {b@\@citeb}{{\bf ?}\@warning
{Citation `\@citeb' on page \thepage \space undefined}}%
      \hbox{\csname b@\@citeb\endcsname}}}{#1}}
\def\citer{\@ifnextchar [{\@tempswatrue\@citexr}{\@tempswafalse\@citexr[]}}
  \def\@citexr[#1]#2{%
    \if@filesw\immediate\write\@auxout{\string\citation{#2}}\fi
    \def\@citea{}\@cite{\@for\@citeb:=#2\do
      {\@citea\def\@citea{--\penalty\@m}\@ifundefined
{b@\@citeb}{{\bf ?}\@warning
{Citation `\@citeb' on page \thepage \space undefined}}%
\hbox{\csname b@\@citeb\endcsname}}}{#1}}

\begin{document}

\begin{titlepage}

\begin{flushright}
{\small
LMU-ASC~29/09\\ 
USTC-ICTS-09-10\\
September 2009
%Draft \today
%hep-ph/yymmnnn
}
\end{flushright}

\vspace{0.5cm}
\begin{center}
{\Large\bf \boldmath
Precision Flavour Physics\\ 
with $B\to K\nu\bar\nu$ and $B\to Kl^+l^-$
\unboldmath}
\end{center}

\vspace{0.5cm}
\begin{center}
{\sc M. Bartsch$^1$, M. Beylich$^1$, G. Buchalla$^1$ and D.-N. Gao$^2$}
\end{center}

\vspace*{0.4cm}

\begin{center}
$^1$Ludwig-Maximilians-Universit\"at M\"unchen, Fakult\"at f\"ur Physik,\\
Arnold Sommerfeld Center for Theoretical Physics, 
D--80333 M\"unchen, Germany\\
\vspace*{0.2cm}
$^2$Interdisciplinary Center for Theoretical Study and Department of
Modern Physics,\\ 
University of Science and Technology of China, Hefei, 
Anhui 230026, China
\end{center}

\vspace{2cm}
\begin{abstract}
\vspace{0.2cm}\noindent
We show that a combined analysis of $B\to K\nu\bar\nu$ and $B\to Kl^+l^-$
allows for new physics tests practically free of form factor uncertainties.
Residual theory errors are at the level of several percent. Our study
underlines the excellent motivation for measuring these modes at
a Super Flavour Factory.
\end{abstract}

\vspace*{3cm}

\vfill
\end{titlepage}

%%%%%%%%%%%%%%%%%%%%%%%%%%%%%%%%%%%%%%%%%%%%%%%%%%%%%%%%%%%%%%%%%
%   Introduction
%%%%%%%%%%%%%%%%%%%%%%%%%%%%%%%%%%%%%%%%%%%%%%%%%%%%%%%%%%%%%%%%%
\section{Introduction}
\label{sec:intro}

The standard model of quark flavour physics has successfully passed all
experimental tests to date. This includes the observation of a substantial
number of rare processes and CP asymmetries, which are consistently 
accounted for within the Cabibbo-Kobayashi-Maskawa (CKM) description of 
quark mixing.
On the other hand, many essential features of the standard model, most
notably in the flavour sector, are still not satisfactorily understood
on a more fundamental level.
Deviations from standard expectations, which could guide us towards a
better understanding, appear to be small in general in view of the
basic agreement between theory and observations.
In this situation precision tests of flavour physics become increasingly
important, which motivates current efforts to build a Super Flavour Factory
\cite{Bona:2007qt,Kageyama:2006zd}. Such a facility will enable an exciting 
program in $B$ physics \cite{Browder:2007gg,Browder:2008em}.

One of the best opportunities in this respect could be provided by the 
study of $b\to s\nu\bar\nu$ transitions, induced by interactions at very
short distances. Theoretically ideal would be an inclusive measurement
of $B\to X_s\nu\bar\nu$, where the hadronic matrix element can be
accurately computed using the heavy-quark expansion. Unfortunately,
because of the missing neutrinos, an inclusive experimental determination
of the decay rate is probably unfeasible.
More promising is the measurement of exclusive channels such as
$B\to K\nu\bar\nu$, $B\to K^*\nu\bar\nu$. In this case a clean theoretical
interpretation requires, however, the control of nonperturbative
hadronic form factors. Direct calculations of form factors suffer from
sizable uncertainties. Additional experimental input to eliminate
nonperturbative quantities can therefore be very useful.
Important examples are the related decays $K\to\pi\nu\bar\nu$, where
the hadronic matrix element can be eliminated with the help of
$K^+\to\pi^0 e^+\nu$ using isospin symmetry.
As discussed in \cite{Buchalla:2000sk}, a similar role could be played
by the semileptonic mode $B\to\pi e\nu$ for the rare decay
$B\to K\nu\bar\nu$. This strategy is limited by the breaking of
$SU(3)$ flavour symmetry of the strong interaction, which is also
difficult to estimate with high accuracy.

In this paper we propose to perform a combined analysis of the
rare decays $B\to K\nu\bar\nu$ and $B\to K l^+l^-$. As we shall discuss,
this option has several advantages for controlling hadronic uncertainties.
It allows us to construct precision observables for testing the
standard model and for investigating new physics effects.
In particular neither isospin nor $SU(3)$ flavour symmetry are required
and form factor uncertainties can be eliminated to a large extent.

The paper is organized as follows. Section 2 summarizes the
experimental status. Section 3 collects basic theoretical results.
It includes a discussion of $B\to K$ form factors, weak annihilation
and nonperturbative corrections in $B\to Kl^+l^-$, and the background
for $B^-\to K^-\nu\bar\nu$ from 
$B^-\to\tau^-\bar\nu_\tau\to K^-\nu_\tau\bar\nu_\tau$.
Precision observables are discussed in section 4. Section 5 comments on 
the effects of new physics and conclusions are presented in section 6.   
Further details on form factors relations and on weak annihilation
are collected in the appendix.

%\afterpage{\clearpage}

%%%%%%%%%%%%%%%%%%%%%%%%%%%%%%%%%%%%%%%%%%%%%%%%%%%%%%%%%%%%%%%%%
%   Experimental status
%%%%%%%%%%%%%%%%%%%%%%%%%%%%%%%%%%%%%%%%%%%%%%%%%%%%%%%%%%%%%%%%%
\section{Experimental status}
\label{sec:expstat}

In this section we summarize briefly the current experimental
situation.
For the branching ratios of the neutrino modes 
$\bar B\to \bar K\nu\bar\nu$ only upper limits are available at present. 
They read
\cite{Amsler:2008zzb,Barberio:2008fa,Chen:2007zk,Aubert:2004ws}
\begin{eqnarray}
B(B^-\to K^-\nu\bar\nu) &<& 14\cdot 10^{-6} \label{kmnunuexp}\\
B(\bar B^0\to\bar K^0\nu\bar\nu) &<& 160\cdot 10^{-6} \label{k0nunuexp}
\end{eqnarray}
Here CP averaged branching fractions are understood.
We note that the limit is more stringent for the $B^-$ channel.

The most accurate experimental results for $B\to Kl^+l^-$ 
are from Belle \cite{Wei:2009zv}. The extrapolated, non-resonant
branching fraction is measured to be
\begin{equation}\label{kllexp}
B(B\to K l^+l^-) = (0.48^{+0.05}_{-0.04}\pm 0.03)\cdot 10^{-6}
\end{equation}
consistent with results from BaBar \cite{Aubert:2006vb}.
The recent paper \cite{Wei:2009zv} also contains information
on the $q^2$-spectrum in terms of partial branching fractions 
for six separate bins. The results for the normalized $q^2$-spectrum,
adapted from \cite{Wei:2009zv}, are given in Table \ref{tab:q2spec}.
%%%%%%%%%%%%%%%%%%%%%%%%%%%%%%%%%%%%%%%%%%%%%%%%%%%%%%%%%%%%%%%%%%%
\begin{table}[t]
\centerline{\parbox{14cm}{\caption{\label{tab:q2spec}
The normalized $q^2$-spectrum for $B\to Kl^+l^-$. 
Shown are the partial branching fractions in six bins of $q^2$ 
(or $s=q^2/m^2_B$) from \cite{Wei:2009zv}, normalized by the central value
of the integrated branching fraction in (\ref{kllexp}).
These quantities are denoted by $\Delta B/B$ in the table.}}}
\begin{center}
\begin{tabular}{*{3}{|c}|}
\hline\hline
$q^2[{\rm GeV}^2]$ & $s$ & $\Delta B/B$ \\
\hline
\hline
0.00--2.00 & 0.00--0.07 & $0.169\pm 0.038$ \\ 
\hline
2.00--4.30 & 0.07--0.15 & $0.096\pm 0.027$ \\
\hline
4.30--8.68 & 0.15--0.31 & $0.208\pm 0.042$ \\
\hline
10.09--12.86 & 0.36--0.46 & $0.115\pm 0.031$ \\
\hline
14.18--16.00 & 0.51--0.57 & $0.079\pm 0.033$ \\
\hline
$> 16.00$ & $> 0.57$ & $0.204\pm 0.042$ \\
\hline
\hline
\end{tabular}
\end{center}
\end{table}
%%%%%%%%%%%%%%%%%%%%%%%%%%%%%%%%%%%%%%%%%%%%%%%%%%%%%%%%%%%%%%%%%%%

%%%%%%%%%%%%%%%%%%%%%%%%%%%%%%%%%%%%%%%%%%%%%%%%%%%%%%%%%%%%%%%%%
%   Basic formulas
%%%%%%%%%%%%%%%%%%%%%%%%%%%%%%%%%%%%%%%%%%%%%%%%%%%%%%%%%%%%%%%%%
\section{\boldmath Theory of $\bar B\to\bar K\nu\bar\nu$ and 
$\bar B\to\bar K l^+l^-$}
\label{sec:basicform}

\subsection{Dilepton-mass spectra and short-distance coefficients}

We define the kinematic quantities
\begin{equation}\label{srkdef}
s=\frac{q^2}{m^2_B} \qquad\quad  r_K=\frac{m^2_K}{m^2_B}
\end{equation}
where $q^2$ is the dilepton invariant mass squared and $m_B$
is the mass of the $B$ meson. 
The kinematical range of $q^2$ and its relation with the
kaon energy $E_K$ are given by
\begin{equation}\label{q2rel}
4 m^2_l\simeq 0\leq q^2\leq (m_B-m_K)^2\qquad\quad
q^2=m^2_B+m^2_K-2 m_B E_K
\end{equation}
We also use the phase-space function
\begin{equation}\label{lkdef}
\lambda_K(s)=1+r^2_K+s^2-2 r_K-2s-2 r_K s
\end{equation}

The differential branching fractions for $\bar B\to\bar K\nu\bar\nu$ and
$\bar B\to\bar Kl^+l^-$ can then be written as follows:
\begin{equation}\label{dbknnds}
\frac{dB(\bar B\to\bar K\nu\bar\nu)}{ds} =
\tau_B\frac{G^2_F\alpha^2m^5_B}{256\pi^5}
|V_{ts}V_{tb}|^2\, \lambda^{3/2}_K(s) f^2_+(s)\, |a(K\nu\nu)|^2
\end{equation}
\begin{equation}\label{dbkllds}
\frac{dB(\bar B\to\bar Kl^+l^-)}{ds} =
\tau_B\frac{G^2_F\alpha^2m^5_B}{1536\pi^5}
|V_{ts}V_{tb}|^2\, \lambda^{3/2}_K(s) f^2_+(s)
\left(|a_9(Kll)|^2 +|a_{10}(Kll)|^2\right)
\end{equation}
Here $\tau_B$ is the $B$-meson lifetime, 
$G_F$ the Fermi constant, $\alpha=1/129$ the electromagnetic coupling
and $V_{ts}$, $V_{tb}$ are elements of the CKM matrix.
A second contribution to the amplitudes proportional to $V^*_{us}V_{ub}$ 
has been neglected. It is below $2\%$ for $\bar B\to\bar Kl^+l^-$ 
and much smaller still for $\bar B\to\bar K\nu\bar\nu$.  

The factorization coefficient $a(K\nu\nu)$ is simply given by
a short-distance Wilson coefficient at the weak scale, $C^\nu_L$,
\cite{Buchalla:2000sk}
\begin{equation}\label{aknn}
a(K\nu\nu)=C^\nu_L=-\frac{1}{\sin^2\Theta_W}\eta_X X_0(x_t)
\end{equation}
where $X_0$ is an Inami-Lim function \cite{Buchalla:1995vs} and
$x_t=m^2_t/M^2_W$, with $m_t=\bar m_t(m_t)$ the $\overline{\rm MS}$
mass of the top quark. The factor $\eta_X=0.994$ accounts for the
effect of ${\cal O}(\alpha_s)$ corrections \cite{Buchalla:1993bv}. 
At this order the residual QCD uncertainty is at the level of $1$-$2\%$ and 
thus practically negligible.

The factorization coefficient $a_9(Kll)$ contains the Wilson
coefficient $\tilde C_9(\mu)$ combined with the short-distance
kernels of the $\bar B\to\bar Kl^+l^-$ matrix elements of four-quark
operators evaluated at $\mu={\cal O}(m_b)$. The coefficient $a_9(Kll)$ 
multiplies the local operator $(\bar sb)_{V-A}(\bar ll)_V$. 
At next-to-leading order (NLO) the result can be extracted from the 
expressions for the inclusive decay $\bar B\to X_sl^+l^-$ 
given in \cite{Buchalla:1995vs,Buras:1994dj,Misiak:1992bc}, 
where also the Wilson coefficients and operators of the
effective Hamiltonian and further details can be found. 
The NLO coefficient reads
\begin{eqnarray}\label{a9kll}
a_9(Kll) &=& \tilde C_9+
   h(z,\hat s)\left(C_1 + 3 C_2+3 C_3+C_4+3C_5+C_6\right)\nonumber\\
&&-\frac{1}{2} h(1,\hat s)\left(4 C_3+4 C_4+3 C_5+C_6\right)\nonumber\\
&&-\frac{1}{2} h(0,\hat s)\left(C_3+3 C_4\right)+
\frac{2}{9}\left(3 C_3+C_4+3C_5+C_6\right) +\frac{2 m_b}{m_B} C_7
\end{eqnarray}
Here
\begin{equation}\label{c9tilde}
\tilde C_9(\mu)=P_0+\frac{Y_0(x_t)}{\sin^2\Theta_W}-
4 Z_0(x_t)+P_E E_0(x_t)
\end{equation}
is the Wilson coefficient in the NDR scheme, $P_0$, $P_E$ are
QCD factors and $E_0$, $Y_0$, $Z_0$ are Inami-Lim functions.
The function $h(z,\hat s)$, $z=m_c/m_b$, $\hat s=q^2/m^2_b$ 
arises from one-loop electromagnetic penguin diagrams, which determine
the matrix elements of four-quark operators. 
In contrast to $\tilde C_9$ the quantity $a_9(Kll)$ is scale and scheme
independent at NLO. To this order the coefficients $C_i$, $i=1,\ldots 7$
in (\ref{a9kll}) are needed only in leading logarithmic approximation (LO).
Note that here the labeling of $C_1$ and $C_2$ is interchanged with respect to
the convention of \cite{Buchalla:1995vs}. 

The coefficient $a_{10}(Kll)$ is 
\begin{equation}\label{a10kll}
a_{10}(Kll)=\tilde C_{10}=-\frac{1}{\sin^2\Theta_W} Y_0(x_t)
\end{equation}

\subsection{Form factors}
\label{subsec:formf}

The long-distance hadronic dynamics of $\bar B\to\bar K\nu\bar\nu$ and
$\bar B\to\bar Kl^+l^-$ is contained in the matrix elements 
\begin{eqnarray}
\langle\bar K(p')|\bar s\gamma^\mu b|\bar B(p)\rangle
&=& f_+(s)\, (p+p')^\mu +[f_0(s)-f_+(s)]\,\frac{m^2_B-m^2_K}{q^2}q^\mu 
\label{fpf0def}\\
\langle\bar K(p')|\bar s\sigma^{\mu\nu}b|\bar B(p)\rangle
&=& i\frac{f_T(s)}{m_B+m_K}\left[(p+p')^\mu q^\nu - q^\mu (p+p')^\nu\right]
\label{ftdef}
\end{eqnarray}
which are parametrized by the form factors $f_+$, $f_0$ and $f_T$.
Here $q=p-p'$ and $s=q^2/m^2_B$. The term proportional to $q^\mu$ in 
(\ref{fpf0def}), and hence $f_0$, drops out when the small lepton masses 
are neglected as has been done in (\ref{dbknnds}) and (\ref{dbkllds}).
The ratio $f_T/f_+$ is independent of unknown hadronic
quantities in the small-$s$ region due to the relations
between form factors that hold in the limit of large kaon energy
\cite{Charles:1998dr,Beneke:2000wa}
\begin{equation}\label{ftfp}
\frac{f_T(s)}{f_+(s)}=\frac{m_B+m_K}{m_B}+{\cal O}(\alpha_s,\Lambda/m_b)
\end{equation}
Here we have kept the kinematical dependence on $m_K$ in the
asymptotic result.
In contrast to $f_+$ the form factor $f_T$ is scale and scheme dependent.
This dependence is of order $\alpha_s$ and has been neglected in (\ref{ftfp}).
Within the approximation we are using we may take $\mu=m_b$ to be
the nominal scale of $f_T$.

We remark that the same result for $f_T/f_+$ is also obtained in the 
opposite limit where the final state kaon is soft, that is in the region 
of large $s={\cal O}(1)$. This follows from the asymptotic expressions
for $f_+$ and $f_T$ in heavy hadron chiral perturbation theory 
\cite{Wise:1992hn,Burdman:1992gh,Falk:1993fr,Casalbuoni:1996pg,Buchalla:1998mt}. 
From this observation we expect (\ref{ftfp}) to be
a reasonable approximation in the entire physical domain. This is indeed 
borne out by a detailed analysis of QCD sum rules on the light cone 
\cite{Ball:2004ye}, which cover a range in $s$ from $0$ to $0.5$.
Relation (\ref{ftfp}) is further discussed in appendix \ref{sec:ftfprel}.

The ratio $f_T/f_+$ enters (\ref{a9kll}) as a prefactor
of $C_7$ from the matrix element of the corresponding
magnetic-moment type operator $Q_7$ \cite{Buchalla:2000sk,Buchalla:1995vs}.  
In writing (\ref{a9kll}) the relation (\ref{ftfp}) has already been used
to eliminate $f_T/f_+$.
Since the $C_7$ term contributes only about $13\%$ to the amplitude
$a_9(Kll)$, the impact of corrections to (\ref{ftfp}) will be
greatly reduced. A $15\%$ uncertainty, which may be expected for the
approximate result (\ref{ftfp}), will only imply an uncertainty of $2\%$ 
for $a_9(Kll)$ or the $\bar B\to\bar K l^+l^-$ differential rate.
In practice, this leaves us with the form factor $f_+(s)$ as the essential
hadronic quantity for both $\bar B\to\bar K\nu\bar\nu$ and 
$\bar B\to\bar K l^+l^-$.

The main emphasis of the present study is on the construction of
clean observables, which are, as far as possible, independent of 
hadronic input. We will therefore consider suitable ratios of
branching fractions where the form factor $f_+(s)$ is eliminated to a 
large extent. In order to assess the residual form factor uncertainties
in these cases, but also to estimate absolute branching fractions,
it will be useful to have an explicit parametrization of
the form factor at hand. 
We employ the parametrization proposed by Becirevic and Kaidalov
\cite{Becirevic:1999kt} in the form
\begin{equation}\label{ffparam}
f_+(s)\equiv f_+(0)\,\frac{1-(b_0+b_1-a_0 b_0)s}{(1-b_0 s)(1-b_1 s)}
= f_+(0)[1+a_0 b_0 \, s +{\cal O}(s^2)]
\end{equation}
The parameter $b_0$ is given by
\begin{equation}\label{b0def}
b_0=\frac{m^2_B}{m^2_{B^*_s}}\approx 0.95 \qquad {\rm for} \quad
m_{B^*_s}=5.41\,{\rm GeV}
\end{equation}
$b_0$ represents the position of the $B^*_s$ pole and will be
treated as fixed, following \cite{Becirevic:1999kt}.
The remaining three parameters $a_0$, $b_1$ and $f_+(0)$ have been
determined from QCD sum rules on the light cone
(LCSR) in \cite{Ball:2004ye}
\begin{equation}\label{a0b1f0}
f_+(0)=0.304\pm 0.042,\qquad a_0\approx 1.5, \qquad b_1=b_0 
\end{equation}
We will treat all three as variable parameters. This also includes
$b_1$, slightly generalizing the expressions from \cite{Ball:2004ye}
where $b_1$ is fixed at $b_0$. 
The value for $f_+(0)$ in (\ref{a0b1f0}) is obtained from the relation
\cite{Ball:2004ye}
\begin{equation}\label{f0ak1}
f_+(0)=0.331\pm 0.041 + 0.25(\alpha_1(1\,{\rm GeV})-0.17)
\end{equation}
using the updated value \cite{Khodjamirian:2004ga,Ball:2005vx}
for the Gegenbauer coefficient $\alpha_1(1\,{\rm GeV})=0.06\pm 0.03$
as quoted in \cite{Buchalla:2008jp}.

With $b_0$ fixed, the parameter $a_0$ introduced
in (\ref{ffparam}) determines the slope of the form factor at small $s$.
We remark that the LCSR method is appropriate for the low-$s$ region,
which will be of particular interest for us.
For completeness we give the relation of our parameters
$f_+(0)$, $a_0$, $b_1$  to the original parameters $c_B$, 
$\alpha\equiv\alpha_B$, $\gamma\equiv\gamma_B$ from \cite{Becirevic:1999kt}:
\begin{equation}\label{bkpar}
f_+(0)=c_B(1-\alpha_B)\, ,\qquad
a_0=\frac{\gamma_B-\alpha_B}{\gamma_B(1-\alpha_B)}\, , \qquad
\frac{b_0}{b_1}=\gamma_B
\end{equation}
As discussed in \cite{Becirevic:1999kt}, the large-energy limit
for the kaon implies the relation $\gamma_B=1/\alpha_B$ or, equivalently, 
$a_0 b_0=b_0+b_1$. 

We determine next our default ranges for the shape parameters
$a_0$ and $b_1$, which will be employed in the subsequent phenomenological
analysis. Three main pieces of information will be used:
The experimental data on the $q^2$ spectrum in Table \ref{tab:q2spec},
the LCSR results in (\ref{a0b1f0}), and asymptotic results for the
form factor at maximum $s$, 
\begin{equation}\label{smax}
s_m=\left(1-\frac{m_K}{m_B}\right)^2
\end{equation}
The third constraint will lead to a relation between $a_0$ and $b_1$.
It follows from the asymptotic expression for $f_+(s_m)$ 
\begin{equation}\label{ffhhchipt}
f_+(s_m)=\frac{g f_B m_B}{2 f_K (m_K+\Delta)}
\quad\qquad \Delta=m_{B^*_s}-m_B
\end{equation}
which can be derived within heavy-hadron chiral perturbation theory
\cite{Wise:1992hn,Burdman:1992gh,Falk:1993fr,Casalbuoni:1996pg,Buchalla:1998mt}.
The largest uncertainty in (\ref{ffhhchipt}) is due to the
$BB^*_s K$ coupling $g$, sometimes also normalized as 
$g_{BB^*_s K}=2 m_B g/f_K$, which is not known precisely.
For the analogous, $SU(3)$ related quantity $g_{BB^*\pi}$
a range of $g_{BB^*\pi}=42\pm 16$ is quoted in \cite{Becirevic:1999kt}.
This corresponds roughly to $g=0.6\pm 0.2$.
In view of this uncertainty, and since the main purpose here is
the estimate of typical numbers, we have neglected subleading 
corrections to (\ref{ffhhchipt}), which may be sizeable \cite{Falk:1993fr}.
We recall that $g$ is of order unity in the large-$m_B$ limit. 

Equating (\ref{ffhhchipt}) with $f_+(s_m)$ from (\ref{ffparam}) and using
\begin{equation}\label{b0sm}
1-b_0 s_m  = \frac{2(m_K+\Delta)}{m_B}
\end{equation}
we obtain
\begin{equation}\label{b1a0c0}
\frac{1-s_m(b_0+b_1-a_0 b_0)}{1-s_m b_1}=\frac{g f_B}{f_+(0) f_K}
\equiv c_0 \approx 2.5\pm 1.0 
\end{equation}
We note that the denominator of the first term in (\ref{b1a0c0}), $1-s_m b_1$, 
scales as $1/m_B$ in the heavy-quark limit, whereas the numerator
remains of order unity (if $a_0$ is not too far below its
typical value of 1.5). The first term then scales as $m_B$,
consistent with the heavy-quark scaling of the second expression. 

The constraint (\ref{b1a0c0}) can also be put in the form
\begin{equation}\label{c0a0b1}
c_0-1=\frac{a_0-1}{\frac{1}{s_m b_0}-\frac{b_1}{b_0}}
\end{equation}
Numerically we have $1/(s_m b_0)=1.279$. Within the uncertainty
of $c_0$, displayed in (\ref{b1a0c0}) above, (\ref{c0a0b1}) implies
a correlation between the shape parameters $a_0$ and $b_1/b_0$.  

\begin{figure}[t]
\begin{center}
%\psfrag{x}[t]{$s$}
%\psfrag{y}[b]{$(dB/ds)/B$}
\resizebox{8cm}{!}{\includegraphics{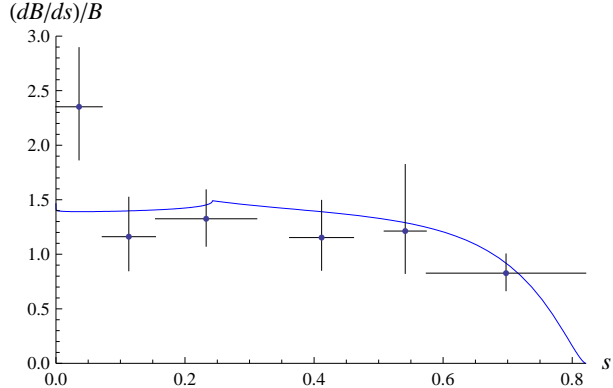}}
\caption{\label{fig:bestfit}
The shape of the $\bar B\to\bar K l^+l^-$ spectrum, $(dB/ds)/B$,
from the Belle data summarized in Table \ref{tab:q2spec} (crosses)
and from theory with the best-fit shape parameters 
$a_0=1.6$, $b_1/b_0=1$ (solid curve).}
\end{center}
\end{figure}

Independently of such theory constraints we might ultimately
want to extract the form factor shape from experimental data. 
In this spirit, we
have investigated how well different values of $(a_0,b_1/b_0)$
fit the current Belle measurements of the dilepton-mass spectrum
in $B\to Kl^+l^-$. For this purpose we show in Table \ref{tab:ffchi2}
%%%%%%%%%%%%%%%%%%%%%%%%%%%%%%%%%%%%%%%%%%%%%%%%%%%%%%%%%%%%%%%%%%%
\begin{table}[t]
\centerline{\parbox{14cm}{\caption{\label{tab:ffchi2}
Values of $\chi^2$ for various combinations of the
form factor shape parameters $a_0$ and $b_1$, determined from
a comparison with the Belle data on the $q^2$ spectrum of
$B\to Kl^+l^-$ (see Table \ref{tab:q2spec}).}}}
\begin{center}
%\begin{tabular}{*{7}{|c}|}
\begin{tabular}{|c||c|c|c|c|c|c|}
\hline\hline
\backslashbox{$b_1/b_0$}{$a_0$} & 
      $1.0$ & $1.2$  & $1.4$ & $1.6$ & $1.8$ & $2.0$ \\
\hline
\hline
$0.5$ & 20.2 & 14.7 & 11.0 & 8.8 & 7.5 & 7.1 \\
\hline
$0.6$ & 20.2 & 14.2 & 10.3 & 8.0 & 6.9 & 6.7 \\
\hline
$0.7$ & 20.2 & 13.5 & 9.4 & 7.1 & 6.2 & 6.4 \\
\hline
$0.8$ & 20.2 & 12.7 & 8.3 & 6.2 & 5.7 & 6.4 \\
\hline
$0.9$ & 20.2 & 11.8 & 7.1 & 5.3 & 5.5 & 7.0 \\
\hline
$1.0$ & 20.2 & 10.5 & 5.7 & 4.7 & 6.2 & 9.2 \\
\hline
\hline
\end{tabular}
\end{center}
\end{table}
%%%%%%%%%%%%%%%%%%%%%%%%%%%%%%%%%%%%%%%%%%%%%%%%%%%%%%%%%%%%%%%%%%%
the $\chi^2$-function
\begin{equation}\label{chi2def}
\chi^2(a_0,b_1)=\sum_{i=1}^6 \frac{(y_i-F_i(a_0,b_1))^2}{\sigma^2_i}
\end{equation}  
where the $y_i$, $i=1,\ldots , 6$, are the experimental values for the 
normalized, partial branching fractions $\Delta B/B$ in each of the six bins,
the $\sigma_i$ are the corresponding errors, and the $F_i$ are the
theoretical expressions depending on $(a_0,b_1)$. 
It is clear that the experimental data are at present not accurate
enough to allow for a precise determination of the form factor shape.
However, the situation should improve in the future. At the moment our 
analysis merely serves to illustrate the general method. Nevertheless,
some regions of parameter space are already disfavoured, in particular
low values of $a_0$, which, as discussed above, is consistent with 
theoretical expectations. We also observe that the best fit is
obtained for $a_0=1.6$, $b_1/b_0=1$.
A comparison of the best-fit shape of the theoretical spectrum with
the Belle data is shown in Fig. \ref{fig:bestfit}

Combining the information above, we adopt the following default 
ranges for the shape parameters
\begin{equation}\label{a0b1range}
1.4\leq a_0\leq 1.8\qquad\quad  0.5\leq b_1/b_0\leq 1.0
\end{equation}
with 
\begin{equation}\label{a0b1ref}
a_0=1.6\qquad\quad  b_1/b_0=1.0
\end{equation}
as our reference values.
Within the range (\ref{a0b1range}) for $a_0$ and $b_1/b_0$
the parameter $c_0$ in (\ref{c0a0b1}) takes values between
$1.5$ and $3.9$, compatible with (\ref{b1a0c0}). 
Our default parameters (\ref{a0b1ref}) are also consistent with
the LCSR results \cite{Ball:2004ye} quoted in (\ref{a0b1f0}).

\subsection{\boldmath $\bar B\to \bar K l^+l^-$: Weak annihilation}
\label{subsec:wa}

As pointed out in \cite{Beneke:2001at}, the exclusive decay
$\bar B\to \bar K l^+l^-$ receives contributions from
weak annihilation diagrams already at leading order in the
heavy-quark limit. In spite of this their impact is numerically small 
because of a strong CKM suppression (for the charged mode) or small 
Wilson coefficients (for the neutral mode).
In this section we will quantify the size of weak annihilation.
Since this contribution is small we work to leading order in $\alpha_s$.

First we consider the case of $\bar B^0\to \bar K^0 l^+l^-$,
and the region of low $s\sim\Lambda/m_b$.
Weak annihilation can then only come from penguin operators giving rise
to the transition $b\bar d\to s\bar d$. A virtual photon emitted from
one of the quarks produces the lepton pair.
The leading annihilation contribution in the heavy-quark limit is generated 
by the gluon-penguin operators $Q_3$ and $Q_4$ \cite{Buchalla:1995vs}.  
This effect can be evaluated using the methods of QCD factorization
\cite{Beneke:2001at,Beneke:2001ev} as discussed in more detail in 
appendix \ref{sec:wabkll}.
The resulting correction to the coefficient $a_9(Kll)$ reads
\begin{equation}\label{da9wa34}
\Delta a_{9,WA,34}=\left(C_4+\frac{1}{3}C_3\right)
\frac{8\pi^2 Q_d f_B f_K}{m_B f_+(s)}\,
\int_0^\infty d\omega\, \frac{\phi_-(\omega)}{\omega-s\, m_B-i\epsilon}
\end{equation}
Here $f_B$, $f_K$ are the meson decay constants and $Q_d=-1/3$ is
the charge of the down-quark in the initial state.
The leading light-cone distribution amplitudes of the $B$ meson
can be expressed by two functions, $\phi_\pm(\omega)$, of which
only $\phi_-$ enters the integral in (\ref{da9wa34}).
For annihilation contributions scale dependent quantities such as the 
Wilson coefficients $C_{3,4}$ will be evaluated at a hard-collinear scale 
of $\mu_h=\sqrt{\mu \Lambda_h}$, where $\mu={\cal O}(m_b)$ and 
$\Lambda_h=0.5\,{\rm GeV}$, following \cite{Beneke:2001ev}.

Our result (\ref{da9wa34}) for weak annihilation agrees with 
eq. (68) of \cite{Beneke:2001at}, once it is adapted to the
case of a pseudoscalar $K$ meson as described immediately after eq. (69)
of \cite{Beneke:2001at}.\footnote{A factor of $(-2 m_b/M_B)$
is missing in front of the annihilation term on the r.h.s. of eq. (41) in
\cite{Beneke:2001at}. We thank Thorsten Feldmann for confirmation.}

For estimates of the annihilation effect we employ the model
functions \cite{Grozin:1996pq}
\begin{equation}\label{phipm}   
\phi_+(\omega)=\frac{\omega}{\omega^2_0}\, e^{-\omega/\omega_0}\qquad\quad
\phi_-(\omega)=\frac{1}{\omega_0}\, e^{-\omega/\omega_0}
\end{equation}
where $\omega_0={\cal O}(\Lambda_{\rm QCD})$ serves to parametrize the
uncertainty related to $\phi_\pm$. A summary of general properties
of these wave functions can be found in \cite{Beneke:2001at,Beneke:2000wa}.
They are satisfied by the parametrizations in (\ref{phipm}).
With $\phi_-(\omega)$ from (\ref{phipm}), and denoting by $\rm Ei(z)$
the exponential integral function, the integral in (\ref{da9wa34})
can be expressed as \cite{Beneke:2001at}
\begin{equation}\label{lambmi}
\lambda^{-1}_{B,-}(s)\equiv 
\int_0^\infty d\omega\, \frac{\phi_-(\omega)}{\omega-s\, m_B-i\epsilon}=
\frac{1}{\omega_0}\, e^{-s m_B/\omega_0}\, [-{\rm Ei}(s m_B/\omega_0)+i\pi]
\end{equation}
At a typical value of $s=0.1$ we have $a_9=3.96 + 0.05 i$ and
$\Delta a_{9,WA,34}=-0.036+0.034 i$.
The correction (\ref{da9wa34}) is seen to reduce the real part of $a_9$ 
by a small amount, which leads to a corresponding reduction of the
branching fraction. This holds if $s\geq 0.37 \omega_0/m_B\approx 0.025$. 
Since the imaginary part of $a_9$ is much smaller than the real part,
its impact on the decay rate is entirely negligible.
Practically it is thus of no consequence that the imaginary part of the 
correction is comparable to the one of $a_9$ and that ${\rm Im}\, a_9$ is 
rather uncertain.
Of particular interest for us is the size of the annihilation effect
on the partially integrated branching fraction.
For $\omega_0=(0.350\pm 0.150)\,{\rm GeV}$, $m_b/2\leq\mu\leq 2m_b$
the reduction of the branching fraction integrated within
$0.03\leq s\leq 0.25$ is at most $1\%$, which is indeed very small.

Weak annihilation contributions to $\bar B^0\to\bar K^0 l^+l^-$
also come from the remaining two QCD penguin operators $Q_5$ and $Q_6$.
Because these have a chiral structure different from $Q_3$, $Q_4$, 
their contribution to weak annihilation is formally suppressed
in $\Lambda/m_b$. It turns out, however, that the suppression
is not very effective numerically in this particular case.
A similar situation is familiar from the factorizing matrix elements
of $Q_5$ and $Q_6$ for $B$ decays into a pair of light pseudoscalar
mesons \cite{Beneke:2001ev}. 
The explicit calculation of the annihilation correction
to $a_9(Kll)$ from $Q_5$ and $Q_6$ proceeds as before and yields 
\begin{equation}\label{da9wa56}
\Delta a_{9,WA,56}=-\left(C_6+\frac{1}{3}C_5\right)
\frac{16\pi^2 Q_d f_B f_K \mu_K}{m^2_B f_+(s)}\,
\int_0^\infty d\omega\, 
\frac{\phi_-(\omega)}{\omega-s\, m_B-i\epsilon}
\end{equation}
In comparison to (\ref{da9wa34}) the correction in (\ref{da9wa56})
carries, apart from the Wilson coefficients, a relative factor of 
$-2\mu_K/m_B\approx -0.75$, where 
$\mu_K=\mu_\pi=m^2_\pi/(m_u+m_d)$ \cite{Beneke:2001ev}.
Here $m_u$, $m_d$ are the $\overline{\rm MS}$ masses evaluated
at the scale $\mu_h$. As anticipated, this relative factor
is not small, although it is of ${\cal O}(\Lambda/m_b)$. 
A typical value for (\ref{da9wa56}), at $s=0.1$, is 
$\Delta a_{9,WA,56}=0.045-0.043 i$. The sign of the real part
is opposite to the case of $\Delta a_{9,WA,34}$ such that there is
a tendency of the two contributions to cancel. Because $|C_6+C_5/3|$ 
is larger than $|C_4+C_3/3|$, it is possible that (\ref{da9wa56})
even dominates over (\ref{da9wa34}). Of course, (\ref{da9wa56}) is
formally a power correction and other power corrections to weak
annihilation do exist. However, the numerically large factor
$2\mu_K/m_B$ is special to the annihilation matrix element of
$Q_6=-2(\bar db)_{S-P}(\bar sd)_{S+P}$ with the presence of 
(pseudo)scalar currents. Taking into account the single power
correction (\ref{da9wa56}) thus appears justified.   
Adding both corrections, (\ref{da9wa34}) and (\ref{da9wa56}),
the net effect on the branching ratio integrated from $s=0.03$ to $0.25$
is a tiny enhancement. This enhancement remains below about $0.3\%$ 
for  $\omega_0=(0.350\pm 0.150)\,{\rm GeV}$ and  $m_b/2\leq\mu\leq 2m_b$.

To summarize, the weak annihilation contributions
to $\bar B^0\to\bar K^0 l^+l^-$, which arise from QCD penguin operators,
are negligibly small in practice, even though they are a leading-power effect.
If the presumably dominant (chirally enhanced) power correction 
(\ref{da9wa56}) is also included, the total impact of weak annihilation  
is further reduced.

In the case of the charged mode $B^-\to K^- l^+l^-$
the weak annihilation terms from QCD penguin operators
are given by the expressions (\ref{da9wa34}) and (\ref{da9wa56})
with the replacement of the quark charge $Q_d\to Q_u$.
Numerically the effect then receives an additional factor of $-2$, which still
yields a negligible correction at the level of about one percent.
Weak annihilation through the tree operators $Q_1$ and $Q_2$,
which exists only for the charged mode, comes with large Wilson
coefficients, but also with a strong Cabibbo suppression.
This correction reads
\begin{equation}\label{da9wa12u}
\Delta a_{9,WA,12u}=-\frac{V^*_{us}V_{ub}}{V^*_{ts}V_{tb}} 
\left(C_1+\frac{1}{3}C_2\right)\frac{8\pi^2 Q_u f_B f_K}{m_B f_+(s)}\,
\int_0^\infty d\omega\, \frac{\phi_-(\omega)}{\omega-s\, m_B-i\epsilon}
\end{equation}
With (\ref{da9wa12u}) the branching ratio integrated from $s=0.03$ to $0.25$
is reduced by less than about $0.6\%$ for  
$\omega_0=(0.350\pm 0.150)\,{\rm GeV}$ and  $m_b/2\leq\mu\leq 2m_b$. 
Taking the three contributions from $Q_{1,2}$, $Q_{3,4}$ and $Q_{5,6}$
together, the reduction remains below $1\%$.
Within an uncertainty of this order, weak annihilation is therefore
negligible for $B^-\to K^- l^+l^-$ as well.

\subsection{\boldmath $\bar B\to \bar K l^+l^-$: Nonperturbative corrections}

In this section we comment on the theoretical framework for
$\bar B\to \bar K l^+l^-$ and on nonperturbative effects beyond those 
that are contained in the form factors.

It is well known that, because of huge backgrounds from 
$\bar B\to\bar K\psi^{(')}\to\bar Kl^+l^-$, the region of $q^2$ containing 
the two narrow charmonium states $\psi=\psi(1S)$ and $\psi'=\psi(2S)$ 
has to be removed by
experimental cuts from the $q^2$ spectrum of $\bar B\to \bar K l^+l^-$.
The overwhelming background from $\psi$ and $\psi'$ is related to a drastic
failure of quark-hadron duality in the narrow-resonance region for the
{\it square\/} of the charm-loop amplitude, as has been discussed
in \cite{Beneke:2009az}. 
Nevertheless, the parts of the $q^2$ spectrum below and above the
narrow-resonance region remain under theoretical control and are sensitive
to the flavour physics at short distances.
A key observation here is that the amplitude is largely dominated by the
semileptonic operators 
\begin{eqnarray}\label{q9q10}
Q_9 &=& (\bar sb)_{V-A}(\bar ll)_V \nonumber\\
Q_{10} &=& (\bar sb)_{V-A}(\bar ll)_A
\end{eqnarray}
which have large coefficients $\tilde C_9$ and $\tilde C_{10}$.
These contributions are perturbatively calculable up to the long-distance
physics contained in the form factor $f_+(s)$. 
The $\bar B\to\bar Kl^+l^-$ matrix elements of four-quark operators,
such as $(\bar sb)_{V-A}(\bar cc)_{V-A}$, are more complicated, but still
systematically calculable. 
Schematically, the $\bar B\to\bar Kl^+l^-$ rate is proportional to
\begin{equation}\label{c94qc10}
|\tilde C_9+\Delta_{4q}|^2 + |\tilde C_ {10}|^2
\end{equation}
where $\Delta_{4q}$ represents contributions from four-quark operators,
for instance charm loops or annihilation effects.
In the present discussion we ignore the small contribution from
$C_7$, which has already been discussed in section \ref{subsec:formf}.
Typical values are $\tilde C_9=4.2$, $\tilde C_{10}=-4.2$ and, for the
charm-loop amplitude 
$\Delta_{4q}\approx (C_1+3 C_2) h(z,\hat s)\approx 0.3$. The last figure
corresponds to an average within $0< s < 0.25$. For large $s$ the charm loop 
develops an imaginary part, but the magnitude is of similar size.
Annihilation effects are negligible as shown in section \ref{subsec:wa}.
Thus $\Delta_{4q}$ is only about a $10\%$ effect, both as a correction to
the $\tilde C_9$ amplitude and to the total rate.
Because this term is numerically subleading the impact of any uncertainties 
in its evaluation will be suppressed. 
We briefly discuss the theory of $\Delta_{4q}$ in the regions of low
and high $q^2$.

In the {\it low-$q^2$ region\/} $\Delta_{4q}$ can be computed using
QCD factorization \cite{Beneke:2001at}.
This approach, which is based on the heavy-quark limit and the large energy
of the recoiling kaon, should work well for the real part of the amplitude
in view of the experience from two-body hadronic $B$ decays 
\cite{Beneke:2007zz} and $B\to K^*\gamma$ \cite{Bosch:2004nd}.
Power corrections of order $\Lambda/m_b\sim 0.1$ in $\Delta_{4q}$ give
only percent level corrections for the differential rate.
The charm loops receive also corrections of order $\Lambda^2/m^2_c$
\cite{Voloshin:1996gw}, which have been estimated at the level of a few 
percent for the exclusive decay $B\to K^*\gamma$ \cite{Khodjamirian:1997tg}.
Since the charm loops are relatively less important in $\bar B\to\bar Kl^+l^-$
by about a factor of five in the rate, the impact of the correction
is reduced. On the other hand, the effect increases somewhat
as $q^2$ approaches the resonance region. In the inclusive case 
$b\to sl^+l^-$ it amounts to a few percent \cite{Buchalla:1997ky},
averaged over the low-$q^2$ region.
We therefore conclude that the $\Lambda^2/m^2_c$ correction 
is unlikely to affect $\bar B\to\bar Kl^+l^-$ in an appreciable way.  

Any quark-level calculation of physical amplitudes involves the
concept of quark-hadron duality. There are no indications that
this assumption, applied to the charm-loops for small $q^2$ up
to about $7\,{\rm GeV}^2$ ($s=0.25$), would introduce an error
in excess of power corrections or perturbative uncertainties
\cite{Beneke:2009az}.

Light-quark loops are generally suppressed by small Wilson coefficients
(QCD penguins) or small CKM factors. Violations of local quark-hadron
duality could come from the presence of light vector resonances
at low $q^2$. To get an order-of-magnitude estimate we consider
the branching ratio of the decay chain $B^-\to K^-\rho^0\to K^-e^+e^-$,
which is measured to be \cite{Amsler:2008zzb}
\begin{eqnarray}\label{bkrholl}
&& B(B^-\to K^-\rho^0)\times B(\rho^0\to e^+e^-) = \nonumber\\
&& (4.2\pm 0.5)\cdot 10^{-6}\, \times\, (4.71\pm 0.05)\cdot 10^{-5}
=(2.0\pm 0.2)\cdot 10^{-10}
\end{eqnarray}
This contribution arises from the $|\Delta_{4q}|^2$ term
in (\ref{c94qc10}) for which quark-hadron duality cannot
be expected to hold \cite{Beneke:2009az}. However, like similar 
contributions with other light vector resonances, it clearly gives
a negligible contribution.  
Resonance effects could be more important in the interference of
$\Delta_{4q}$ with $\tilde C_9$ in (\ref{c94qc10}). In this case
they are part of the hadronic amplitude that is dual to light-quark
loops in the partonic calculation. Oscillations in $s$ due to
light resonances are bound to be small because of the smallness of
the light-quark loops. Integration over the low-$q^2$ region will
further reduce such violations of local duality to a negligible level.    

In the {\it high-$q^2$ region\/} the appropriate theoretical
framework for the computation of $\Delta_{4q}$ is an
operator product expansion exploiting the presence of the large
scale $q^2\sim m^2_b$. 
This concept has been used in \cite{Buchalla:1998mt} in analyzing
the endpoint region of $b\to sl^+l^-$, which is governed by few-body 
exclusive modes. A systematic treatment, including the discussion
of subleading corrections, has been given in \cite{Grinstein:2004vb}. 
Power corrections are generally smaller than for low $q^2$.
Terms of order $\Lambda/m_b$ arise at order $\alpha_s$ \cite{Grinstein:2004vb}
and the analogue of the $\Lambda^2/m^2_c$ corrections at small $q^2$ 
now contribute only at order $\Lambda^2/m^2_b$ \cite{Buchalla:1997ky}.
More important are perturbative corrections to the leading-power term, 
which however can be systematically improved.
Finally, uncertainties could come from violations of local quark-hadron 
duality. By duality violation we mean deviations of the OPE
calculation, at fixed $q^2$ and in principle including all perturbative
and power corrections, from the real-world hadronic result.
Such violations are related in particular to oscillations of $\Delta_{4q}$
in $s$ due to higher charmonium resonances. These oscillations are absent
in the smooth OPE result. 
To first order in $\Delta_{4q}$ only its real part contributes to 
(\ref{c94qc10}). Implementing the higher charmonium resonances,
$\psi(3770)$, $\psi(4040)$, $\psi(4160)$, $\psi(4415)$, and the
hadronic $c\bar c$ continuum in the approximation of Kr\"uger and Sehgal
\cite{Kruger:1996cv}, we estimate the relative amplitude of oscillations
in $\tilde C_9 + {\rm Re}\,\Delta_{4q}$ to be of order $10$ to $20\%$.
We expect these local variations to be averaged out when the spectrum
is integrated over $s$ \cite{Buchalla:1998mt} such that the residual
uncertainty will be somewhat reduced.
The $s$-integration, which is also phenomenologically motivated,
effectively produces a smearing that leads to a more `globally'
defined quantity where duality is better fulfilled.
As discussed in \cite{Beneke:2009az},
global duality in this sense cannot be expected to hold for the
second order term $|\Delta_{4q}|^2$ in (\ref{c94qc10}).
On the other hand, this contribution is numerically very small,
at the level of few percent, and duality violations will only have
a minor effect. To illustrate this point we consider the decay
chain $B^-\to K^-\psi(3770)\to K^-e^+e^-$, which can be viewed
as a resonance contribution to $|\Delta_{4q}|^2$. In the
case of the narrow charmonium states a similar contribution leads
to the very large resonance background mentioned above. 
Here we have \cite{Amsler:2008zzb} 
\begin{eqnarray}\label{bkpsill}
&& B(B^-\to K^-\psi(3770))\times B(\psi(3770)\to e^+e^-) = \nonumber\\
&& (4.9\pm 1.3)\cdot 10^{-4}\, \times\, (9.7\pm 0.7)\cdot 10^{-6}
=(4.8\pm 1.3)\cdot 10^{-9}
\end{eqnarray}
This indicates that resonance contributions are rather small, in
agreement with our previous remarks. 
In conclusion, we have argued that duality violations from the resonance
region at high $q^2$ are at a moderate level and should not spoil
a precision of theoretical predictions for (partially) integrated
branching ratios of $\bar B\to\bar Kl^+l^-$ at the level of several percent.
A more detailed investigation of this issue would be of interest and
will be given elsewhere.

In the present analysis we ignore higher order electroweak and
QED radiative corrections. The latter could modify the decay modes and 
their ratios presumably at the level of several percent. 
The leading effects could be taken into account if it should be
required by the experimental precision.

\subsection{\boldmath $B^-\to K^-\nu\bar\nu$: Background from
$B^-\to\tau^-\bar\nu_\tau\to K^-\nu_\tau\bar\nu_\tau$}
\label{subsec:nunubkgr}

The decay  $B^-\to\tau^-\bar\nu_\tau$ followed by $\tau^-\to K^-\nu_\tau$
produces a background for the short-distance reaction $B^-\to K^-\nu\bar\nu$, 
which has been discussed recently in \cite{Kamenik:2009kc}.
The branching fractions of $B^-\to\tau^-\bar\nu_\tau$ and 
$\tau^-\to K^-\nu_\tau$ are given by
\begin{eqnarray}\label{bbtaunu}
B(B^-\to\tau^-\bar\nu_\tau) &=& \tau_B\frac{G^2_F m_B m^2_\tau f^2_B}{8\pi}
|V_{ub}|^2 \,\left(1-\frac{m^2_\tau}{m^2_B}\right)^2 \nonumber\\ 
&=& 0.87\cdot 10^{-4}\left(\frac{f_B}{0.2\,{\rm GeV}}\right)^2
\left(\frac{|V_{ub}|}{0.0035}\right)^2
\end{eqnarray}
\begin{equation}\label{btauknu}
B(\tau^-\to K^-\nu_\tau)=\tau_\tau\frac{G^2_F m^3_\tau f^2_K}{16\pi}
|V_{us}|^2 \,\left(1-\frac{m^2_K}{m^2_\tau}\right)^2 = 7.46\cdot 10^{-3}
\end{equation}
The numerical values are based on the input parameters in
Table \ref{tab:parinput}. 
\begin{table}[t]
\centering
\begin{tabular}[h]{|c|c|c|c|c|}
  \hline\hline
   $M_W$\big[GeV\big] & $\bar m_t(\bar m_t)$\big[GeV\big] & 
   $m_B$\big[GeV\big] & 
   $m_K$\big[GeV\big] & $m_{B_s^{*}}$\big[GeV\big] \\
  \hline
  $80.4$ & $166$  & $5.28$  & $0.496$ & $5.41$   \\
  \hline\hline
   $\bar m_b$\big[GeV\big]& $\bar m_c$\big[GeV\big]&$\alpha$ & 
   $\sin^2{\theta_W}$ & $\vert V^*_{ts}V_{tb}\vert$ \\
  \hline
   $4.2$ & $1.3$ &$1/129$& $0.23$ & $0.039$  \\
  \hline\hline
   $f_+(0)$ & $f_B$ \big[GeV\big] &  $f_K$ \big[GeV\big] &
  $\Lambda_{\overline{\rm MS},5}$\big[GeV\big] & 
   $\tau_{B^+}$ ($\tau_{B^0}$)\big[ps\big]\\
  \hline
   $0.304\pm0.042$ & $0.2$ & $0.16$ & $0.225$ & $1.64$ ($1.53$)\\
  \hline\hline
%$G_F^2 \big[{\rm GeV}^{-5}\,{\rm ps}^{-1}\big]$  $206.52$
\end{tabular}
\caption{\label{tab:parinput} Input parameters.}
\end{table}
The background from the decay chain
$B^-\to\tau^-\bar\nu_\tau\to K^-\nu_\tau\bar\nu_\tau$
gives a contribution to the dilepton-mass spectrum, which can
be written as
\begin{equation}\label{bknnbkgr}
\frac{dB(B^-\to K^-\nu_\tau\bar\nu_\tau)_{bkgr}}{ds}=
B(B^-\to\tau^-\bar\nu_\tau)\, B(\tau^-\to K^-\nu_\tau)\,
\frac{2t((1-t)(t-r_K)-ts)}{(1-t)^2 (t-r_K)^2}
\end{equation}
where we used (\ref{srkdef}) and $t\equiv m^2_\tau/m^2_B$. 
The spectrum in (\ref{bknnbkgr}) extends from $s=0$ to
$s=(1-t)(1-r_K/t)=0.818$. The maximum $s$ happens to coincide almost 
exactly with the endpoint of the spectrum in the short-distance decay 
$B^-\to K^-\nu\bar\nu$, $s_m=0.821$ \cite{Kamenik:2009kc}.
Integrated over the full range in $s$, the phase-space factor
in (\ref{bknnbkgr}) gives $1$. This reproduces the result for
$B(B^-\to K^-\nu_\tau\bar\nu_\tau)_{bkgr}$ in the narrow-width
approximation for the intermediate $\tau$ lepton.

If the decay sequence $B^-\to\tau^-\bar\nu_\tau\to K^-\nu_\tau\bar\nu_\tau$
cannot be distinguished experimentally from the short-distance decay
$B^-\to K^-\nu\bar\nu$, this background should be subtracted from the
measured rate of $B^-\to K^- +\,{\rm ``invisible"}$ to obtain the true
short-distance branching fraction. 
In the standard model we find from (\ref{bbtaunu}) and (\ref{btauknu})
\begin{equation}\label{bkgrsm}
B(B^-\to K^-\nu_\tau\bar\nu_\tau)_{bkgr} = (0.65\pm 0.16)\cdot 10^{-6}
\end{equation} 
assuming an uncertainty of $25\%$ due to $f_B$ and $|V_{ub}|$.
The central value amounts to about $15\%$ of the short-distance branching 
fraction (\ref{bnnnum}).
A subtraction of (\ref{bkgrsm}) from the measured branching ratio
would then lead to an uncertainty of about $0.16/4.4=4\%$
on $B(B^-\to K^-\nu\bar\nu)$.
This error might be further reduced in the future with improved
determinations of $f_B$ and $|V_{ub}|$.

Probably the best method to control the background is to
use the experimental measurement of $B(B^-\to\tau^-\bar\nu_\tau)$.
In this way, any new physics component in the latter decay will
automatically be removed from the measurement of $B^-\to K^-\nu\bar\nu$. 
This will simplify the new physics interpretation of
the measured $B(B^-\to K^-\nu\bar\nu)$.
The present experimental value for $B(B^-\to\tau^-\bar\nu_\tau)$
is \cite{Barberio:2008fa,Aubert:2007xj,Adachi:2008ch}
\begin{equation}\label{btaunuexp}
B(B^-\to\tau^-\bar\nu_\tau)_{exp}=(1.43\pm 0.37)\cdot 10^{-4}
\end{equation}
Together with (\ref{btauknu}) this gives
\begin{equation}\label{bkgrexp}
B(B^-\to K^-\nu_\tau\bar\nu_\tau)_{bkgr} = (1.1\pm 0.28)\cdot 10^{-6}
\end{equation} 
The $26\%$ uncertainty in (\ref{btaunuexp}) thus implies an error of
about $6\%$ in $B(B^-\to K^-\nu\bar\nu)$, assuming the central value
of (\ref{bnnnum}). However, by the time when $B(B^-\to K^-\nu\bar\nu)$
will be measured at a Super Flavour Factory, $B(B^-\to\tau^-\bar\nu_\tau)$
will be simultaneously known with high precision.
According to \cite{Browder:2007gg,Browder:2008em} the expected accuracy
is about $4\%$ or better. Assuming the central values of 
(\ref{btaunuexp}) and (\ref{bnnnum}) as before, the background subtraction
will then lead to an error of only about $1\%$ in $B(B^-\to K^-\nu\bar\nu)$.

We conclude that the background from 
$B^-\to\tau^-\bar\nu_\tau\to K^-\nu_\tau\bar\nu_\tau$
pointed out in \cite{Kamenik:2009kc} has to be taken into account 
for a precise measurement of the short-distance branching fraction 
$B(B^-\to K^-\nu\bar\nu)$. It needs to be subtracted from the
experimental signal, but this should ultimately be possible with
essentially negligible uncertainty. The background discussed
here is absent in the case of the neutral mode 
$\bar B^0\to \bar K^0 \nu\bar\nu$.

%%%%%%%%%%%%%%%%%%%%%%%%%%%%%%%%%%%%%%%%%%%%%%%%%%%%%%%%%%%%%%%%%
%   Precision observables
%%%%%%%%%%%%%%%%%%%%%%%%%%%%%%%%%%%%%%%%%%%%%%%%%%%%%%%%%%%%%%%%%
\section{Precision observables}
\label{sec:precision}

\subsection{Theory expectations for branching fractions}

The input parameters we will use in the present analysis are collected 
in Table \ref{tab:parinput}.

To begin our discussion of numerial results we consider first the
integrated branching ratios of $B^-\to K^-\nu\bar\nu$ and $B^-\to K^-l^+l^-$.
For the neutrino mode we find
\begin{equation}\label{bnnnum}
B(B^-\to K^-\nu\bar\nu)\cdot 10^6 =4.4\, ^{+1.3}_{-1.1}\, (f_+(0))
\,\, ^{+0.8}_{-0.7}\, (a_0)\,\, ^{+0.0}_{-0.7}\, (b_1)
\end{equation}
We have displayed the sensitivity to the form factor parameters,
which are by far the dominant sources of uncertainty.
The form factor normalization $f_+(0)$ has the largest impact,
while the shape parameters are relatively less important.

The fully integrated, non-resonant $\bar B\to\bar K l^+l^-$
branching fraction can be evaluated in a similar way.
This quantity corresponds essentially to the experimental
result in (\ref{kllexp}), which has been obtained by cutting
out the large background from the two narrow charmonium resonances
and by extrapolating the measurements to the entire $q^2$ range to
recover the total non-resonant rate. 
The precise correspondence between theoretical and experimental
results will depend on the details of the cuts and the extrapolation
procedure. We will treat the resonance region more carefully
later when we study precision observables.
For our present discussion we simply identify the integral
over the non-resonant spectrum in (\ref{dbkllds}) with
the measurement in (\ref{kllexp}). 
This appears justified as the error from this identification
is expected to be below the experimental uncertainty. 
Adopting these considerations we compute 
\begin{equation}\label{bllnum}
B(B^-\to K^-l^+l^-)\cdot 10^6 =0.58\, ^{+0.17}_{-0.15}\, (f_+(0))
\,\, ^{+0.10}_{-0.09}\, (a_0)\,\, ^{+0.00}_{-0.09}\, (b_1)
\,\, ^{+0.04}_{-0.03}\, (\mu)
\end{equation}
In addition to the still dominant dependence on the form factor we
have in this case a non-negligible perturbative uncertainty,
which we estimate in the standard way through a variation of
the scale $\mu$ between $m_b/2$ and $2 m_b$ around the reference
value of $\mu=m_b$. The scale dependence is at a rather moderate
level of $\pm 6\%$ with NLO accuracy, much smaller than the
error from the hadronic parameters.
Within sizeable, mainly theoretical uncertainties,
the prediction (\ref{bllnum}) is in agreement with the measurement
in (\ref{kllexp}).

\begin{figure}[t]
\begin{center}
\psfrag{x}[t]{$s$}
\psfrag{y}[b]{$(dB/ds)/B$}
\resizebox{8cm}{!}{\includegraphics{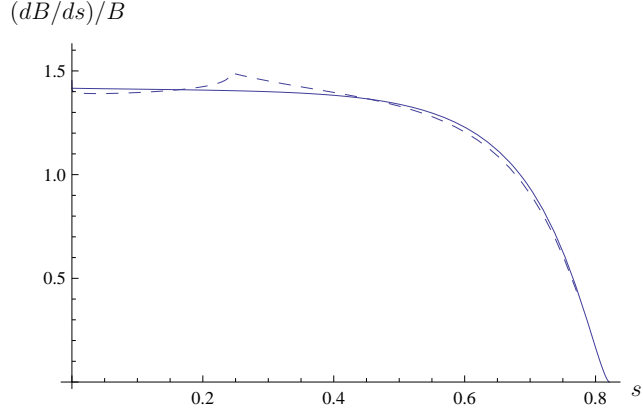}}
\caption{\label{fig:bkllnn}
The dilepton invariant-mass spectra for $\bar B\to\bar K\nu\bar\nu$ (solid) 
and $\bar B\to\bar K l^+l^-$ (dashed). For easier comparison
of the shapes the plotted differential branching fractions,
$dB/ds$ versus $s=q^2/m^2_B$, were normalized by their integral $B$. 
The reference values have been used for all parameters.}
\end{center}
\end{figure}

Whereas the individual branching fractions (\ref{bnnnum}) and
(\ref{bllnum}) suffer from large hadronic uncertainties, we expect their
ratio to be under much better theoretical control. It is obvious that the 
form factor normalization $f_+(0)$ cancels in this ratio. 
Moreover, as illustrated in Fig. \ref{fig:bkllnn},
the shape of the $q^2$ spectrum is almost identical for the
two modes. This is because the additional $q^2$-dependence
from charm loops in $B\to Kl^+l^-$, compared to $B\to K\nu\bar\nu$,
is numerically only a small effect outside the region of the
narrow charmonium states.
As a consequence, also the dependence on the form factor shape
will be greatly reduced in the ratio
\begin{equation}\label{rdef}
R=\frac{B(B^-\to K^-\nu\bar\nu)}{B(B^-\to K^-l^+l^-)}
\end{equation}
Numerically we find
\begin{equation}\label{rnum}
R=7.59\, ^{+0.01}_{-0.01}\, (a_0)\,\, ^{+0.00}_{-0.02}\, (b_1)
\,\, ^{-0.48}_{+0.41}\, (\mu)
\end{equation}
This prediction is independent of form factor uncertainties
for all practical purposes. It is limited essentially by
the perturbative uncertainty at NLO of $\pm 6\%$. 
Using the experimental result in (\ref{kllexp}), the theory
prediction (\ref{rnum}), and assuming the validity of the
standard model, we obtain
\begin{equation}\label{knnrkll}
B(B^-\to K^-\nu\bar\nu)=R\cdot B(B^-\to K^-l^+l^-)_{exp}=
(3.64\pm 0.47)\cdot 10^{-6}
\end{equation}
With an accuracy of $\pm 13\%$, limited at present by
the experimental error, this result is currently the most
precise estimate of $B(B^-\to K^-\nu\bar\nu)$.

Since isospin breaking effects in the decay rates are very small, 
the branching ratios for the decays $\bar B^0\to\bar K^0\nu\bar\nu$
and $\bar B^0\to\bar K^0 l^+l^-$ are given by the branching
ratios for the corresponding $B^-$ modes multiplied by a factor
of $\tau(\bar B^0)/\tau(B^-)=0.93$.

\subsection{Precision observables: ratios of branching fractions}

In order to obtain theoretically clean observables, the
region of the two narrow charmonium resonances $\psi(1S)$ and
$\psi(2S)$ has to be removed from the $q^2$ spectrum of $B\to Kl^+l^-$. 
This leaves two regions of interest, the low-$s$ region below
the resonances, and the high-$s$ region above.
For the present analysis we define these ranges as
\begin{equation}\label{slohi}
\begin{array}{rl}
{\rm low}\ s: \qquad\quad & 0\leq s \leq 0.25\\
{\rm high}\ s: \qquad\quad & 0.6\leq s \leq s_m
\end{array}
\end{equation}
The resonance region $0.25 < s < 0.6$ corresponds to
the $q^2$ range $7\,{\rm GeV}^2 < q^2 < 16.7\,{\rm GeV}^2$.
For our standard parameter set the total rate for
$B\to K\nu\bar\nu$ or $B\to Kl^+l^-$ (non-resonant)
is divided among the three regions, low-$s$, narrow-resonance, high-$s$,
as $35 : 48 : 17$.

We first concentrate on the low-$s$ region, where $B^-\to K^- l^+l^-$
can be reliably calculated. To ensure an optimal cancellation of the
form factor dependence, one may restrict also the neutrino mode
to the same range in $s$ and define  
\begin{equation}\label{r25def}
R_{25}\equiv \frac{\int_0^{0.25}ds\ dB(B^-\to K^-\nu\bar\nu)/ds}{
\int_0^{0.25}ds\ dB(B^-\to K^- l^+l^-)/ds}
\end{equation}
This ratio is determined by theory to very high precision.
Displaying the sensitivity to the shape parameters and the
renormalization scale one finds
\begin{equation}\label{r25num}
R_{25}=7.60\, ^{-0.00}_{+0.00}\, (a_0)\,\, ^{-0.00}_{+0.00}\, (b_1)
\,\, ^{-0.43}_{+0.36}\, (\mu)
\end{equation}
The form factor dependence is seen to cancel almost perfectly in $R_{25}$.
The shape pa\-ra\-me\-ters affect this quantity at a level of only 0.5 per 
mille. One is therefore left with the perturbative uncertainty, estimated 
here at about $\pm 5\%$ at NLO. 

The independence of any form factor uncertainties in $R_{25}$ comes at 
the price of using only $35\%$ of the full $B^-\to K^-\nu\bar\nu$ rate.
We therefore consider a different ratio, which is defined by 
\begin{equation}\label{r256def}
R_{256}\equiv \frac{\int_0^{s_m}ds\, dB(B^-\to K^-\nu\bar\nu)/ds}{
\int_0^{0.25}ds\, dB(B^-\to K^- l^+l^-)/ds +
\int_{0.6}^{s_m}ds\, dB(B^-\to K^- l^+l^-)/ds}
\end{equation}
In this ratio the fully integrated rate of 
$B^-\to K^-\nu\bar\nu$ is divided by the integrated rate
of $B^-\to K^- l^+l^-$ with only the narrow-resonance region removed.
This ensures use of the maximal statistics in both channels.
Due to the missing region in $B^-\to K^- l^+l^-$ the dependence on the
form factor shape will no longer be eliminated completely,
but we still expect a reduced dependence.  
Numerically we obtain, using the same input as before,
\begin{equation}\label{r256num}
R_{256}=14.60\, ^{+0.28}_{-0.38}\, (a_0)\,\, ^{+0.10}_{-0.02}\, (b_1)
\,\, ^{-0.80}_{+0.62}\, (\mu)
\end{equation}
This estimate shows that the uncertainty from $a_0$ and $b_1$ is indeed
very small, at a level of about $\pm 3\%$. With better empirical
information on the shape of the spectrum this could be further improved.

We conclude that ratios such as those in (\ref{r25def}) and (\ref{r256def}), 
or similar quantities with modified cuts, are theoretically very well under
control. They are therefore ideally suited for testing the standard model 
with high precision.

\subsection{Precision observables: $\bar B\to\bar Kl^+l^-$ with lattice input}
\label{subsec:lattice}

Until now our strategy has been to achieve accurate predictions
by eliminating the form factor dependence altogether.
A variant of our analysis consists in taking a single hadronic
parameter, the form factor at one particular value of $q^2$,
$f_+(s_0)$, $s_0=q^2_0/m^2_B$, as additional theory input.
The shape of the form factor can be fitted to the experimental
spectrum as discussed in sec. \ref{subsec:formf}.
At the expense of one extra hadronic parameter it is then possible to 
probe short distance physics based on $\bar B\to\bar Kl^+l^-$ alone. 
The necessary input $f_+(s_0)$ could come from lattice QCD
calculations.

This approach is analogous to the method pursued in
\cite{Becher:2005bg} to determine $|V_{ub}|$ from
$B\to\pi l\nu$. In this case lattice results on the
$B\to\pi$ form factor at a typical value of $q^2=16\,{\rm GeV}^2$
were considered as theory input. Experimental data on the spectrum and
decay rate of $B\to\pi l\nu$ can then be used to extract $|V_{ub}|$.
In order to describe the form factor shape, \cite{Becher:2005bg}  
employed dispersive bounds and a related class of general
parametrizations \cite{Boyd:1997qw,Boyd:1994tt,Bourrely:1980gp}.
These more sophisticated parametrizations may also be applied
in our case, if more than two shape parameters should be
required to fit the data with the appropriate precision.
For the time being the form factor parametrization used
here is completely sufficient. As shown in \cite{Becher:2005bg},
the limiting factor is the value of $f_+(s_0)$.
We remark that in the case of $\bar B\to\bar Kl^+l^-$, the
narrow-resonance region should be removed from the analysis
when performing the fit to the form factor shape. 

To illustrate the method we extract the form
factor $f_+(s_0)$ at point $s_0=16\,{\rm GeV}^2/m^2_B$
from the measurement in (\ref{kllexp}).
Using the best-fit shape parameters $a_0=1.6$ and $b_1/b_0=1$
we obtain
\begin{equation}\label{fplfit}
f_+(s_0) = 1.05 \pm 0.06\, ({\rm BR}) \pm 0.03\, (\mu)
\end{equation} 
where the first error is from the measured branching ratio
in (\ref{kllexp}) and the second error is from scale dependence.
The sensitivity to the shape parameters is comparable to the
uncertainty from the branching ratio.

Unlike the parameters $a_0$ and $b_1$, the value of $f_+(s_0)$
determined in this way, assuming the standard model, is sensitive to the 
normalization of the branching ratio and therefore to new physics effects.
These could be detected through a comparison with QCD calculations of 
$f_+(s_0)$.  Our default choice of hadronic parameters leads to
$f_+(s_0)=1.16$ but the uncertainty is larger than $15\%$.
Within the coming five to ten years a precision of
$\pm 4\%$ might be achieved for the form factor
$f_+(s_0)$ in lattice QCD \cite{Buchalla:2008jp}.

%%%%%%%%%%%%%%%%%%%%%%%%%%%%%%%%%%%%%%%%%%%%%%%%%%%%%%%%%%%%%%%%%
%   New physics
%%%%%%%%%%%%%%%%%%%%%%%%%%%%%%%%%%%%%%%%%%%%%%%%%%%%%%%%%%%%%%%%%
\section{New physics}
\label{sec:newphysics}

The branching fractions of $B\to K\nu\bar\nu$ and $B\to Kl^+l^-$ are
sensitive to physics beyond the standard model.
If the new physics would modify both of them by (almost) the same factor, 
this change would not be visible in the ratios $R_{25}$ or $R_{256}$
discussed in sec. \ref{sec:precision}. In that case, the new physics
could still be seen by studying $B\to Kl^+l^-$ separately with the method
described in sec. \ref{subsec:lattice}. An example is a scenario with
modified $Z$-penguin contributions \cite{Buchalla:2000sk} interfering
constructively with the standard model terms.
In this case $R_{25}$ is changed only by a small amount.

In general, however, nonstandard dynamics will have a different impact on
$B\to K\nu\bar\nu$ and $B\to Kl^+l^-$. The excellent theoretical control
over the ratios $R_{25}$ or $R_{256}$ will help to reveal even moderate
deviations from standard model expectations. 

One example is the scenario with modified $Z$-penguin contributions 
\cite{Buchalla:2000sk} mentioned before, if these contributions
interfere destructively with those of the standard model.
In that case the ratios $R_{25}$ or $R_{256}$ could be
significantly suppressed. The modified $Z$-penguin scenario
may be realized, for instance, in supersymmetric models
\cite{Buchalla:2000sk,Altmannshofer:2009ma}.

Another class of theories that do change the ratios are those where
$B\to Kl^+l^-$ remains standard model like while $B\to K\nu\bar\nu$
receives an enhancement (or a suppression). Substantial enhancements of
$B(B\to K\nu\bar\nu)$ are still allowed by experiment, in fact much more
than for $B\to Kl^+l^-$.

A first example are scenarios with light invisible scalars $S$
contributing to $B\to KSS$ \cite{Altmannshofer:2009ma}.
This channel adds to $B\to K\nu\bar\nu$, which is measured as
$B\to K + {\rm invisible}$. If the scalars have nonzero mass,
$B\to KSS$ could be distinguished from $B\to K\nu\bar\nu$ through the
missing-mass spectrum. On the other hand, if the mass of $S$ is small,
or the resolution of the spectrum is not good enough, a discrimination
of the channels may be difficult. The corresponding increase in
$B(B\to K\nu\bar\nu)$ could be cleanly identified through the ratios
$R_{25}$ and $R_{256}$.

A second example is given by topcolor assisted technicolor 
\cite{Buchalla:1995dp}. A typical scenario involves new strong dynamics,
together with extra $Z'$ bosons, which distinguishes the third generation
from the remaining two. The resulting flavour-changing neutral currents 
at tree level may then predominantly lead to transitions between 
third-generation fermions such as $b\to s\nu_\tau\bar\nu_\tau$.
An enhancement of $B(B\to K\nu\bar\nu)$ would result and might
in principle saturate the experimental bound (\ref{kmnunuexp}).
An enhancement of $20\%$, which should still be detectable, would probe 
a $Z'$-boson mass of typically $M_{Z'}\approx 3\,{\rm TeV}$.
A similar pattern of enhanced $B\to K\nu\bar\nu$ and SM like
$B\to Kl^+l^-$ is also possible in generic $Z'$ models
\cite{Altmannshofer:2009ma}. 

A more detailed exploration of new physics in $B\to K\nu\bar\nu$
and $B\to K l^+l^-$ is beyond the scope of this article.
The examples mentioned above illustrate that the ratios of branching
fractions considered here exhibit a significant sensitivity
to interesting new physics scenarios. 
The subject of new physics in $b\to s\nu\bar\nu$ transitions has
been discussed in \cite{Grossman:1995gt} and most recently
in \cite{Altmannshofer:2009ma}. New physics in
$B\to K l^+l^-$ has been studied in \cite{Bobeth:2007dw},
including the information from angular distributions.

%%%%%%%%%%%%%%%%%%%%%%%%%%%%%%%%%%%%%%%%%%%%%%%%%%%%%%%%%%%%%%%%%
%     Conclusions
%%%%%%%%%%%%%%%%%%%%%%%%%%%%%%%%%%%%%%%%%%%%%%%%%%%%%%%%%%%%%%%%%
\section{Conclusions}
\label{sec:conclusion}
In this paper we have studied precision tests of the standard model
through a combined analysis of $B\to K\nu\bar\nu$ and $B\to Kl^+l^-$.
The main points can be summarized as follows:
\begin{itemize}
\item
After removing the narrow-resonance region the branching fraction
of $B\to Kl^+l^-$ can be reliably computed. 
The dominant amplitude from semileptonic operators is simply a calculable
expression times the form factor $f_+$. 
QCD factorization for low $q^2$ and OPE for high $q^2$ allow one to 
treat also matrix elements of 4-quark operators in a systematic way.
These are dominated by charm loops, which are numerically small 
contributions to begin with. Since the tensor form factor $f_T$ can be
related to $f_+$ in the heavy-quark limit, the entire $B\to Kl^+l^-$
amplitude becomes calculable in terms of practically a single
hadronic quantity, the form factor $f_+(s)$.
\item
The decay mode $B\to K\nu\bar\nu$ is a particularly clean process. It is
completely determined by short-distance physics at the weak scale 
up to the same form factor $f_+(s)$. For the charged mode
the background due to $B^-\to\tau^-\bar\nu_\tau\to K^-\nu_\tau\bar\nu_\tau$
should be subtracted from the experimental signal, but this
will be possible without introducing any appreciable uncertainty.
\item
The form factor uncertainty can be eliminated by constructing suitable
ratios of (partially) integrated rates such as $R_{25}$ in (\ref{r25def})
and $R_{256}$ in (\ref{r256def}). The resulting quantities can be computed
with high accuracy. The cancellation of form factors is exact and
does not require the use of approximate flavour symmetries.
\item
The perturbative uncertainty of the ratios is estimated to be
$\pm 5\%$ at next-to-leading order (NLO). This can be further improved
by a NNLO analysis along the lines of \cite{Beneke:2001at}.
Uncertainties from other sources are at the level of several percent.
Some refinements in controlling them should still be possible.
\item
Based on the current measurements of $B^-\to K^-l^+l^-$ we
predict 
\begin{equation}
B(B^-\to K^-\nu\bar\nu)=(3.64\pm 0.47)\cdot 10^{-6}\, ,
\end{equation}
at present the most accurate estimate of this quantity.
\item
The ratios of $B\to K\nu\bar\nu$ and $B\to Kl^+l^-$ rates have
an interesting sensitivity to new physics. The new physics
reach benefits from the high accuracy of the standard model
predictions.
A complementary new physics test is possible based on $B\to Kl^+l^-$
alone, exploiting experimental information on the $q^2$ spectrum,
if $f_+$ at one particular value of $q^2$ is used as input from lattice QCD.  
\item
New physics in the Wilson coefficients factorizes
from low-energy hadronic physics in a simple way. The latter is
essentially contained only in $f_+$. Our analysis can thus
be generalized to specific new physics scenarios in a straightforward 
manner.
\end{itemize}
Our proposal puts $B\to K\nu\bar\nu$ as a new physics probe in the same class
as $K\to\pi\nu\bar\nu$, the `golden modes' of kaon physics.
$B\to K\nu\bar\nu$ together with $B\to Kl^+l^-$ thus hold
exciting opportunities for $B$ physics at a Super Flavour Factory.

%%%%%%%%%%%%%%%%%%%%%%%%%%%%%%%%%%%%%%%%%%%%%%%%%%%%%%%%%%%%%%%%%
%     Appendix
%%%%%%%%%%%%%%%%%%%%%%%%%%%%%%%%%%%%%%%%%%%%%%%%%%%%%%%%%%%%%%%%%
\appendix

\section{\boldmath Relation between form factors $f_T$ and $f_+$}
\label{sec:ftfprel}

The three form factors in (\ref{fpf0def}) and (\ref{ftdef}),
$f_T(s)$, $f_+(s)$, and $f_0(s)$ or equivalently
\begin{equation}\label{fmdef}
f_-(s)\equiv \left[f_0(s)-f_+(s)\right]\frac{m^2_B-m^2_K}{q^2}
\end{equation} 
are related in the heavy-quark limit. If we multiply (\ref{fpf0def}) 
and (\ref{ftdef}) by $v_\mu\equiv p_\mu/m_B$ and use the equation of 
motion for the heavy quark, $\not\! v b=b$, we find
\begin{equation}\label{ftfpfm}
\frac{2 m_B}{m_B+m_K}f_T=f_+-f_-
\end{equation}
Similarly, multiplying (\ref{fpf0def}) with $v_\mu$, using $\not\! v b=b$,
and comparing the result with $q_\mu\cdot$ (\ref{fpf0def}), where the
quark equations of motion are used on the left-hand side, one finds
\begin{equation}\label{fpfm}
f_+=-f_-
\end{equation} 
Together (\ref{ftfpfm}) and (\ref{fpfm}) imply
\begin{equation}\label{ftfphql}
\frac{f_T(s)}{f_+(s)}=\frac{m_B+m_K}{m_B}
\end{equation}
The relations (\ref{ftfpfm}) -- (\ref{ftfphql}) have been obtained in 
\cite{Isgur:1990kf} in the heavy-quark limit. They apply immediately
to the case where the kaon is soft, since then the heavy-quark mass
is the only large energy scale in the problem. On the other hand,
the derivation makes no explicit reference to the kaon energy and
it has been argued in \cite{Isgur:1990jg} that these relations should
be valid in the entire kinematic domain.
This conjecture can be justified within the framework of
soft-collinear effective theory (SCET) \cite{Bauer:2000yr,Bauer:2001yt},  
using the form factor
relations in the large recoil limit \cite{Charles:1998dr,Beneke:2000wa},
which also lead to (\ref{ftfpfm}) -- (\ref{ftfphql}).
A related discussion can be found in \cite{Hill:2005ju}.

We may thus use (\ref{ftfphql}) in the entire range of
$q^2$ between $0$ and $(m_B-m_K)^2$. Since this expression for
$f_T/f_+$ is an asymptotic result in the heavy-quark limit 
$m_b\gg\Lambda_{\rm QCD}$, an important issue is the question
of subleading terms. These can be power corrections
in $\Lambda_{\rm QCD}/m_B$ and perturbative QCD corrections.   
The perturbative corrections were computed in the heavy-quark and
large recoil energy limit in \cite{Beneke:2000wa} with the result
(in the NDR scheme with $\overline{\rm MS}$ subtraction)
\begin{eqnarray}\label{ftfpas}
\frac{m_B}{m_B+m_K}\frac{f_T(s)}{f_+(s)}=
1 &-& \frac{\alpha_s(\mu) C_F}{4\pi}\left[\ln\frac{\mu^2}{m^2_b} +
\frac{4 E_K}{m_B-2 E_K}\ln\frac{2E_K}{m_B}\right] \nonumber \\
&-& \frac{\alpha_s(\mu_h) C_F}{4\pi}\frac{4\pi^2 f_B f_K}{N f_+(s)E_K\lambda_B}
\int_0^1du\,\frac{\phi_K(u)}{1-u}
\end{eqnarray}
where $\mu={\cal O}(m_b)$, $\mu_h={\cal O}(\sqrt{\Lambda m_b})$, 
and $E_K={\cal O}(m_b)$ depends on $s=q^2/m^2_B$ through (\ref{q2rel}).
We note that this relation remains valid when the kaon is soft, 
with $E_K={\cal O}(\Lambda)$. In that case expression (\ref{ftfpas})
simplifies. The term with $\alpha_s(\mu_h)$ is no longer perturbative
since the effective scale $\mu_h$ becomes soft. However, the entire term
is power suppressed $\sim\Lambda/m_b$ because $f_B\sim1/\sqrt{m_b}$
and $f_+(1)\sim\sqrt{m_b}$. The second term in square brackets
is also power suppressed and only the first correction 
$\sim\alpha_s\ln(\mu/m_b)$ survives.
The $\alpha_s$ corrections can be consistently taken into account at NNLO 
even though at present, for low $q^2$, the second term from hard
spectator interactions still introduces an uncertainty of about 
$5$ -- $10\%$.

Power corrections to the heavy-quark limit are more difficult to
compute. An estimate can be obtained from light-cone QCD sum rules
\cite{Ball:2004ye}, which indicate that $f_T/f_+$ deviates from 
$1+m_K/m_B$ by less than $\pm 5\%$ for $0 < s < 0.5$.
The sum rule calculations include $\alpha_s$ corrections within
their framework.

From these considerations we conclude that the relation (\ref{ftfphql})
should be correct to within $\pm 10\%$.

\section{\boldmath Weak annihilation in $\bar B\to\bar K l^+l^-$}
\label{sec:wabkll}

Weak annihilation contributes to $\bar B^0\to \bar K^0 l^+l^-$
through QCD penguin operators. These induce the transition
$b\bar d\to s\bar d$ where the valence quarks $b\bar d$ of the $\bar B^0$
meson are annihilated and transformed into the constituents of the final 
state $\bar K^0$. The virtual photon producing the lepton pair may
be emitted from any of the four quarks in this transition.
In a similar manner the process $b\bar u\to s\bar u$ gives rise to weak
annihilation in $B^-\to K^-l^+l^-$, where the transition comes from
QCD penguins and from doubly Cabibbo suppressed tree operators.  

To be specific we treat the case of $\bar B^0\to \bar K^0 l^+l^-$ first.
Here the leading-power contribution to weak annihilation comes from
the QCD penguin operators
\begin{eqnarray}\label{q3q4}
Q_3 &=& (\bar d_ib_j)_{V-A}(\bar s_jd_i)_{V-A} + \ldots \nonumber\\
Q_4 &=& (\bar db)_{V-A}(\bar sd)_{V-A} + \ldots
\end{eqnarray} 
The ellipsis refers to similar terms with $d$ replaced by $u$, $c$, $s$
and $b$, which do not contribute at the order we are considering.
Colour indices are denoted by $i$, $j$. 

The kinematics of the annihilation process is conveniently described
in terms of two lightlike four-vectors $n_\pm$. Their components
can be chosen, without loss of generality, as
\begin{equation}\label{npmdef}
n^\mu_\pm=(1,0,0,\pm 1)
\end{equation}
The momenta of the $B$ meson, the kaon and the lepton pair,
$p$, $k$ and $q$, respectively, can then be written as
\begin{equation}\label{pkqnpm}
p=\frac{m_B}{2}(n_++n_-),\qquad\quad
k=\frac{m^2_B-q^2}{2m_B}n_+,\qquad\quad
q=\frac{q^2}{2m_B} n_+ +\frac{m_B}{2}n_-
\end{equation}
For now we assume that the dilepton mass $q^2$ counts as order 
$\Lambda m_b$,
appropriate for the low-$q^2$ region. This means that momentum $q$ is
nearly lightlike and approximately in the direction of $n_-$, whereas $k$  
is lightlike and has the direction of $n_+$ if we neglect the kaon mass. 

Consider next the $b\bar d\to s\bar d$ annihilation diagram
where the virtual photon is emitted from the $\bar d$ in the
initial state. We will denote by $A_{d1}$ the contribution of this
diagram to the $\bar B^0\to \bar K^0 l^+l^-$ matrix element of $Q_4$.
Viewed as a function of the $\bar d$ four-momentum $l={\cal O}(\Lambda)$,
the diagram has the form
\begin{equation}\label{flplus}
F(l)=F^{(0)}(l_+)+l^\mu_\perp\, F^{(1)}_\mu(l_+)
\end{equation}
up to terms with a relative power suppression in $\Lambda/m_b$.
The momentum $l$ is decomposed into light-cone coordinates
$l_\pm=n_\mp\cdot l$ and $l_\perp$, $n_\pm\cdot l_\perp=0$,
with respect to the two vectors $n_\pm$ in (\ref{npmdef}).
The expression for the light-cone projector of the $B$ meson in momentum
space, appropriate for an amplitude of the type shown in (\ref{flplus}),
has been derived in \cite{Beneke:2000wa}. It is given by
\begin{equation}\label{bdbarl} 
b\bar d \equiv i\frac{f_B m_B}{4}\frac{1+\not\! v}{2}
\left[\phi_+(\omega) \not\! n_+ +\phi_-(\omega)\left(\not\! n_- -\omega
\gamma^\nu_\perp\frac{\partial}{\partial l^\nu_\perp}\right)\right]\gamma_5
\end{equation}
The derivative in (\ref{bdbarl}) extracts the $F^{(1)}$ contribution
in (\ref{flplus}). After it has been applied, $l_\perp$ has to be set
to zero and $l_+$ is identified with $\omega$.

Calculating the contribution $A_{d1}$ with the projector in (\ref{bdbarl})
we find
\begin{equation}\label{ad1}
A_{d1}=e^2 f_B f_K\, (\bar u\!\! \not\! k v)
\left[\frac{2 Q_{d1}}{m_B}\int^\infty_0 d\omega\,
\frac{\phi_-(\omega)}{\omega - q^2/m_B} + \frac{Q_{d1}}{q^2}\right]
\end{equation}
Here $\bar u\gamma_\mu v$ is the lepton current in momentum space
and $Q_{d1}=-1/3$ the down-quark charge.
The second term in (\ref{ad1}) is of the same order in $\Lambda/m_b$
as the first term. It has a pole in $q^2$ and is inconsistent with
electromagnetic gauge invariance. However, it is structure independent
(it depends only on $f_B$, $f_K$, not on the distribution amplitudes)
and is cancelled by corresponding contributions from the remaining
three diagrams. These diagrams, where the photon is emitted from
the $b$ quark, the $s$ quark and the final-state $\bar d$ quark ($d2$)
give explicitly
\begin{equation}\label{absd2}
A_b+A_s+A_{d2}=-e^2 f_B f_K\, (\bar u\!\! \not\! k v)
\frac{Q_b-Q_s+Q_{d2}}{q^2}
\end{equation}
as contributions to the matrix element of $Q_4$.
Charge conservation implies that $Q_b-Q_s+Q_{d2}\equiv Q_{d1}$,
which guarantees the cancellation of the $1/q^2$ term in (\ref{ad1})
by (\ref{absd2}). The first term in (\ref{ad1}) then leads to the
result (\ref{da9wa34}) once the matrix element of $Q_3$ is included,
which is the same as the one of $Q_4$ up to a factor $1/3$ from colour. 

We remark that the $1/q^2$ term and half of the first term in (\ref{ad1}) 
come from the $F^{(1)}$ part in (\ref{flplus}). 
The latter contribution, as well as all $1/q^2$ terms are absent at
leading power for weak annihilation in $B\to K^*\gamma$
\cite{Beneke:2001at,Bosch:2001gv}, in contrast to the present case.  

The result in (\ref{da9wa34}) develops a logarithmic singularity when
$q^2$ becomes soft of order $\Lambda^2$, corresponding to the formal limit
$q^2\to 0$. Since the singularity is integrable, the soft region
is power suppressed in comparison to the branching ratio
integrated from $0$ to $q^2\sim \Lambda m_b$, as has been discussed
in \cite{Beneke:2001at}.

The annihilation contribution in (\ref{da9wa56}) from operators 
$Q_5=-2(\bar d_ib_j)_{S-P}(\bar s_jd_i)_{S+P}+\ldots$ and 
$Q_6=-2(\bar db)_{S-P}(\bar sd)_{S+P}+\ldots$ 
is obtained in analogy to the case of $Q_3$, $Q_4$. Also for
$Q_5$, $Q_6$ there are structure-independent terms $\sim 1/q^2$,
which cancel when all four diagrams are added. The remaining result
is again due to the diagram where the photon is emitted from
the $\bar d$ quark in the initial state.

%%%%%%%%%%%%%%%%%%%%%%%%%%%%%%%%%%%%%%%%%%%%%%%%%%%%%%%%%%%%%%%%%
%     Acknowledgements
%%%%%%%%%%%%%%%%%%%%%%%%%%%%%%%%%%%%%%%%%%%%%%%%%%%%%%%%%%%%%%%%%
\section*{Acknowledgements}
We thank Andrzej Buras and Thorsten Feldmann for discussions.
This work was supported in part by the DFG cluster of excellence
`Origin and Structure of the Universe' and by the DFG
Graduiertenkolleg GK 1054.
D.N.G is supported in part by the NSF of China under grant No. 10775124 
and by the Scientific Research Foundation for the Returned
Overseas Chinese Scholars, State Education Ministry.

\vspace*{1cm}

%%%%%%%%%%%%%%%%%%%%%%%%%%%%%%%%%%%%%%%%%%%%%%%%%%%%%%%%%%%%%%%%%
%     References
%%%%%%%%%%%%%%%%%%%%%%%%%%%%%%%%%%%%%%%%%%%%%%%%%%%%%%%%%%%%%%%%%

\end{document}